\documentclass[9pt,twocolumn,twoside]{opticajnl}

\journal{opticajournal} 
\setboolean{shortarticle}{false}
\usepackage{dblfloatfix}

\title{Optimal input excitations for suppressing nonlinear instabilities in multimode fibers}

\author[1,*]{Kabish Wisal}
\author[2]{Chun-Wei Chen}
\author[2]{Zeyu Kuang}
\author[2]{Owen D. Miller}
\author[2]{Hui Cao}
\author[2]{A. Douglas Stone}

\affil[1]{Department of Physics, Yale University, New Haven, CT 06520, USA}
\affil[2]{Department of Applied Physics, Yale University, New Haven, CT 06520, USA}
\affil[*]{Corresponding author: kabish.wisal@yale.edu}

\begin{abstract}
Wavefront shaping has become a powerful tool for manipulating light propagation in various complex media undergoing linear scattering. Controlling nonlinear optical interactions with spatial degrees of freedom is a relatively recent but growing area of research. A wavefront-shaping-based approach can be used to suppress nonlinear stimulated Brillouin scattering (SBS) and transverse mode instability (TMI), which are the two main limitations to power scaling in high-power narrowband fiber amplifiers.  Here we formulate both SBS and TMI suppression as optimization problems with respect to coherent multimode input excitation in a given multimode fiber. We develop an efficient method for finding the globally optimal input excitation for SBS and TMI suppression using linear programming. We theoretically show that optimally exciting a standard multimode fiber leads to roughly an order of magnitude enhancement in output power limited by SBS and TMI, compared to fundamental-mode-only excitation. We find that the optimal mode content is robust to small perturbations and our approach works even in the presence of mode dependent loss and gain. Optimal mode content can be excited in real experiments using spatial light modulators, creating a novel platform for instability-free ultrahigh-power fiber lasers.

\end{abstract}

\begin{document}
\maketitle
\section{Introduction}

Coherent input wavefront shaping allows controlling light propagation in complex media undergoing linear scattering, enabling focusing, light delivery, and energy deposition~\cite{mosk2012controlling,ploschner2015seeing,rotter2017light,gigan2022roadmap,cao2022shaping,cao2023controlling}. The ability and limits of control and optimization of various properties via wavefront shaping in linear scattering is relatively straightforward to calculate and understand~\cite{feng1988correlations,beenakker1997random,goetschy2013filtering}.  In many cases (e.g., focusing, light--matter interactions) it reduces to finding the extremal eigenvalues of some linear scattering operator~\cite{kim2012maximal,ambichl2017focusing,bouchet2021maximum} and it is often possible to measure the relevant scattering operator experimentally~\cite{popoff2010measuring,yu2013measuring,carpenter2016complete,xiong2016spatiotemporal}.  However, the ability to control and optimize nonlinear optical scattering~\cite{boyd2008nonlinear,agrawal2000nonlinear,agrawal2011nonlinear,wright2022physics}, which plays an important role in a myriad of applications such as creating new light sources~\cite{dudley2006supercontinuum,otterstrom2018silicon}, ultrafast optics~\cite{maine1988generation,weiner2011ultrafast}, optical computing~\cite{zuo2019all,wright2022deep,wang2022optical,wang2023image,pai2023experimentally} and imaging~\cite{pantazis2010second,min2011coherent,ravasi2013nonlinear,katz2014noninvasive}, is much less amenable to rigorous theoretical analysis. Of particular interest are certain nonlinear interactions, which can lead to destructive instabilities in beam propagation, such as transverse mode instability (TMI)~\cite{Cesar2020transverse,eidam2011experimental,jauregui2011impact,jauregui2012physical,smith2011mode,Hansen2013theoretical,dong2013stimulated,zervas2019transverse}, stimulated Brillouin scattering (SBS)~\cite{atmanspacher1987stimulated,kobyakov2010stimulated,rakich2012giant,wolff2015stimulated,wolff2021brillouin,merklein2022100}, and modulation instability~\cite{agrawal1989modulation,guasoni2015generalized,wright2016self,dupiol2017intermodal}. 

\begin{figure*}[b]
    \centering
    \includegraphics[width=0.9\textwidth]{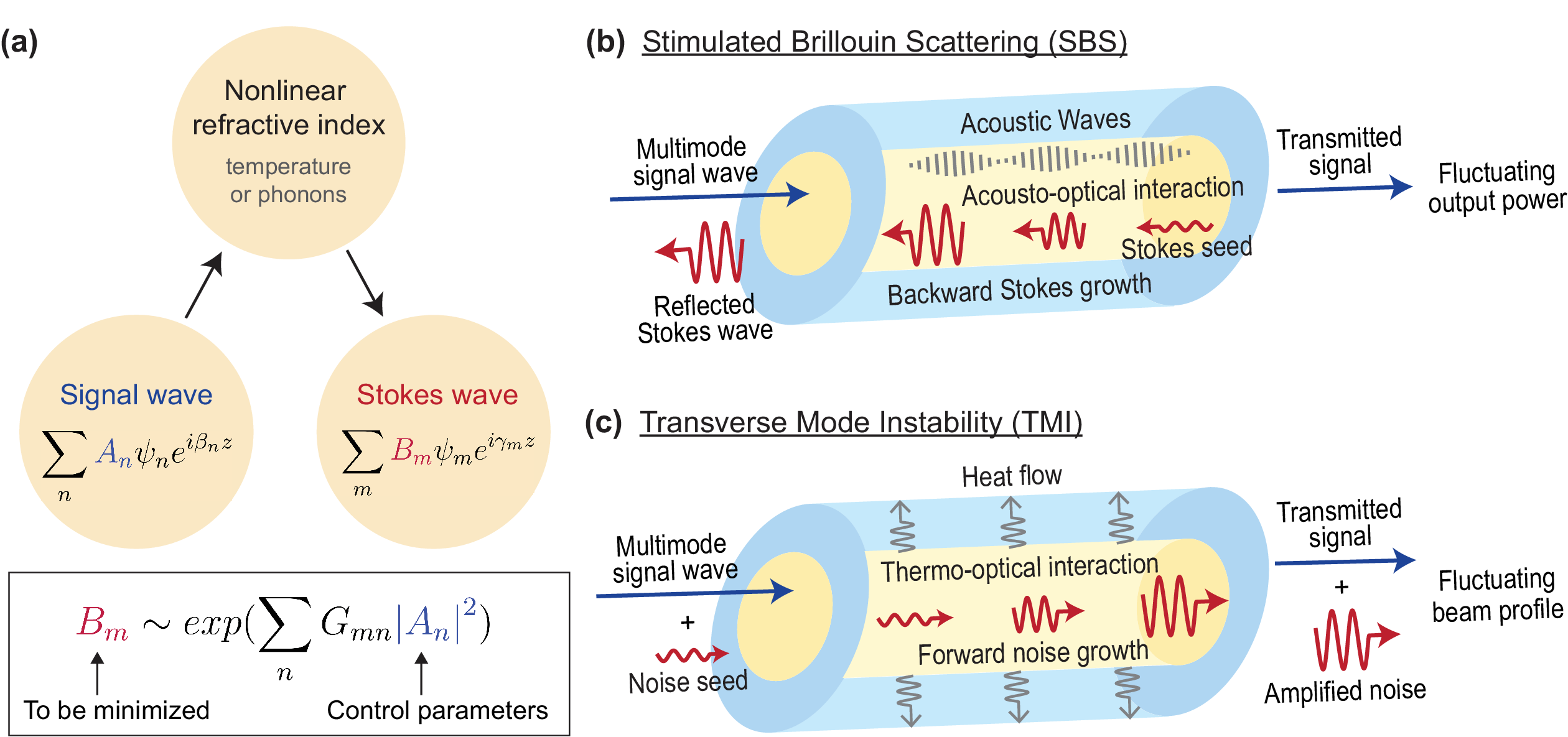}
    \caption{(a) Overview of multimode input optimization approach for suppressing nonlinear instabilities in fibers. The signal wave launched into the fiber generates nonlinear refractive index change. This causes nonlinear scattering light causing exponential growth of noise at specific frequencies, which upon significant growth produces instability. These instabilities can be suppressed by minimizing the growth rate of noise power by controlling the distribution of signal power in various modes at the input. (b) An important nonlinear instability of this kind is SBS, which involves backward reflection of light at a down-shifted frequency by the acoustic phonons. For a large enough signal power, almost all the light is reflected, creating an extremely low transmission. Both the transmitted and reflected power also fluctuate in time which can damage the upstream equipment.(c) Another important nonlinear instability is TMI which results from the growth of noise in various modes due to thermo-optical scattering. As signal undergoes amplification it generates heat (due to the quantum defect) which flows into the cladding and creates temperature fluctuation causing dynamic refractive index variations. The resulting optical scattering causes growth in noise in the forward direction which interferes with the signal producing a fluctuating beam profile at the output. Both TMI and SBS can be suppressed by optimizing the input excitation as described in (a).  }
    \label{fig:Fig1}
\end{figure*}

In contrast to the case of linear scattering, the possibility of using control of the spatial degrees of freedom of the input fields for manipulating nonlinear optical phenomena has been relatively little studied until recently, although interest is now growing~\cite{frostig2017focusing,tzang2018adaptive,deliancourt2019wavefront, ni2023nonlinear, moon2023measuring, yao2012controlling}. A platform of particular relevance is multimode optical fibers where a number of instabilities and non-linear processes can enable or limit various applications~\cite{wright2015controllable,wright2022physics,tzang2018adaptive,chen2023mitigating,chen2023suppressing, ferraro2024calorimetry}. In the past few years 
spatial wavefront shaping has been used in passive fibers for suppressing SBS~\cite{chen2023mitigating}, enhancing stimulated Raman scattering and four-wave mixing due to Kerr nonlinearity~\cite{tzang2018adaptive}, and for demonstrating focusing through a fiber amplifier with nonlinear gain saturation and thermal effects~\cite{florentin2017shaping,chen2024exploiting}. Conversely, nonlinear mode coupling due to the Kerr nonlinearity has been utilized for spatial self beam cleaning~\cite{krupa2017spatial,deliancourt2019wavefront}. In some of these cases the degree of control and optimization was determined empirically via feedback and optimization of some cost function~\cite{chen2023mitigating,tzang2018adaptive,florentin2017shaping}; in others theory allowed calculation of the relevant target function for a given input wavefront, but didn't predict the theoretical optimum~\cite{chen2023suppressing,wisal2023theorySBS,wisal2024theoryTMI}.  In the current work we show that a realistic model for certain practically relevant nonlinear instabilities in multimode fiber (MMF) can generate a tractable optimization problem for mitigating those instabilities, allowing us to find the global optimum over all possible input wavefronts for a key physical parameter of interest, the power threshold for instabilities.
 
The nonlinear instabilities we study here involve degradation of a narrow-band signal in a MMF via nonlinear scattering that alters the signal and transfers signal energy to undesired modes at lower frequencies (see Fig.~\ref{fig:Fig1}a). Two important instabilities of this kind are SBS and TMI, whose origins are described in detail in the next paragraph. Generically, a multimode signal wave is sent into a passive or an active fiber, which induces dynamic nonlinear refractive-index changes. This results in nonlinear scattering which creates light at new frequencies via exponential growth of noise in various transverse modes in the forward or backward direction~\cite{kobyakov2010stimulated,Hansen2013theoretical,dong2013stimulated}. The growth rate depends linearly on the signal power in various modes. Above a certain signal power, defined as the instability threshold, the noise power becomes a significant fraction of the signal power, leading to a depleted transmission (for counter-propagating noise, Fig.~\ref{fig:Fig1}b) or a fluctuating beam profile (for co-propagating noise, Fig.~\ref{fig:Fig1}c). While the noise grows exponentially at negative frequency shifts for any strength of the nonlinearity, the instability threshold can be maximized by finding the optimal wavefront which minimizes the noise growth rate in the fastest growing mode. We show that the resulting optimization problem can be mapped to a linear-programming problem ~\cite{rigler2005optimization,karloff2008linear} and a globally optimal wavefront can be obtained with standard computationally efficient optimization techniques, for any given MMF. As mentioned above, finding the global optimum for wavefront shaping in media with nonlinear interactions is typically very challenging. By mapping the nonlinear instability growth to a linear program in the input parameters, we are able to find the global optimum for suppressing these instabilities.

Our approach for suppressing instabilities can be highly useful in high-power fiber amplifiers, which provide the most promising platform for generating ultra-high laser power~\cite{nilsson2011high,Cesar2013high,Zervas2013high,dong2016fiber}. The power scaling in these fiber amplifiers is primarily limited by SBS and TMI. 
Using MMFs and wavefront shaping for suppressing these instabilities offers a novel strategy for instability-free high power operation, which could enable several new technologies, including improved gravitational-wave detection~\cite{LIGO2012}, advanced manufacturing~\cite{Welding2018}, and directed energy~\cite{Defense2019}. 

SBS is a result of nonlinear scattering of light by acoustic phonons generated by optical forces. A schematic of SBS in a MMF is shown in Fig.~\ref{fig:Fig1}b. A signal wave is launched in the fiber, which creates optical forces in the medium, giving rise to acoustic phonons. These phonons scatter the signal wave in the backward direction causing exponential growth in the reflected wave (seeded by noise) at a down-shifted frequency. The growth rate of the reflected power in each transverse mode depends linearly on the signal power in various modes. Above a certain signal power, defined as the SBS threshold, most of the signal power is reflected back, creating a significant loss in transmitted power and limiting the total output power. Significant research efforts have been made to suppress SBS (or equivalently increase the SBS threshold) in \emph{single-mode} fibers~\cite{yoshizawa1993stimulated,kobyakov2005design,liu2007suppressing,limpert2012yb,supradeepa2013stimulated,hawkins2021kilowatt}. However suppressing SBS while maintaining a narrow laser linewidth~\cite{fu2017review}, as is needed in many key applications~\cite{loftus2007spectrally,buikema2019narrow}, remains a challenge. In work involving several of the current authors~\cite{chen2023mitigating,chen2022suppressing}, it was recently shown that coherent selective mode excitation in passive \emph{multimode} fibers can be used to obtain a substantially higher (a factor of $\sim$3.5) SBS threshold, compared to single-mode excitation, while maintaining narrow linewidth. This was in good agreement with  the theoretical model we are using here~\cite{wisal2023theorySBS}. However in this previous work no analytic or numerical method was presented to find the \emph{globally optimal} input excitation. Here, we provide a theoretical method to find the globally optimal input excitation of modes for obtaining the maximum SBS threshold in any given MMF, within the model which agrees with the previous experiments~\cite{chen2023mitigating}. We find that a 9.6$\times$ higher SBS-threshold can be achieved upon optimal excitation in a standard MMF with typical parameters, compared to the fundamental mode (FM)-only excitation. The optimal mode selection minimizes the peak of the Brillouin gain spectrum while maximally broadening its linewidth, without affecting the signal linewidth. The SBS suppression provided by this mode content is also robust to perturbations. 

Our optimization approach can similarly be applied to suppress TMI, which involves a fluctuating output beam profile caused by nonlinear thermo-optical scattering and is primarily present in active fibers~\cite{Cesar2020transverse}. A schematic of TMI in a fiber amplifier is shown in Fig.~\ref{fig:Fig1}c. As a signal wave
undergoes amplification, it generates heat (due to quantum defects) which flows into the cladding and creates temperature fluctuations causing dynamic refractive index variations. The resulting nonlinear optical scattering causes exponential growth of noise in the forward direction
at a rate depending linearly on the signal power in each transverse mode. For a high enough signal power, defined as the TMI threshold, the power in the noise at the output becomes significant, and it interferes with the signal producing a fluctuating output-beam profile.  Several efforts have been made to suppress TMI and increase the TMI threshold utilizing nearly-single-mode fibers~\cite{Otto2013controlling,Robin2014modal,Smith2013increasing,jauregui2013passive,tao2015mitigating,Li2017experimental,Cesar2018pump,zhang2019bending}. Most of these approaches strive to maintain a single-mode operation, thus avoiding a fluctuating beam profile. However, this is extremely challenging as the quantum-defect heating increases the core refractive index leading to the propagation of additional modes even in a nominally-single-mode fiber. Recently, it was shown theoretically~\cite{wisal2024theoryTMI} and numerically~\cite{chen2023suppressing} by several of the current authors that the TMI threshold can be significantly increased by equally exciting multiple modes in a highly multimode fiber. Equal excitation is much more efficient in suppressing TMI than SBS in active fibers, due to the distinct physical mechanisms at play: thermo-optical interactions in TMI versus acousto-optical interactions in SBS~\cite{wisal2024theoryTMI,wisal2023theorySBS}.  However, while equal excitation is quite effective in suppressing TMI, it is not the optimal input excitation. Below we will find the optimal input mode content for TMI suppression in a typical MMF and show that it produces a 17$\times$ higher TMI threshold compared to FM-only excitation, which is significantly higher even compared to that achieved by equal mode excitation (13$\times$ the TMI threshold under FM-only excitation). Our approach for finding the optimal wavefront works even when we include non-idealities present in real MMF amplifiers, such as mode-dependent loss (MDL) or mode-dependent gain.

Selective mode excitation can and has been implemented via spatial light modulators in experiments~\cite{chen2023mitigating, gomes2022near}. The optimal solution is expected to differ from the theoretical predictions, due to limits on experimental control of the input modal superposition, or experimental non-idealities~\cite{ho2011statistics} not accounted for in the theoretical model. In such a case, the optimal wavefront would need to be obtained through search algorithms~\cite{vellekoop2015feedback,wu2019thorough,tzang2018adaptive}. Nonetheless, the optimal enhancement to SBS and TMI threshold obtained with our model would provide the upper bound for what can be achieved by multimode excitation. Additionally, the optimal mode content provides insight into the physical mechanism behind both SBS and TMI suppression that can be utilized to improve the search algorithms. Finally, our method provides a novel application of linear optimization theory to a nonlinear optical phenomenon, which addresses the important problem of improving power scaling in high-power fiber lasers.

\section{Optimal SBS suppression}

\begin{figure*}[t!]
    \centering
    \includegraphics[width=0.9\textwidth]{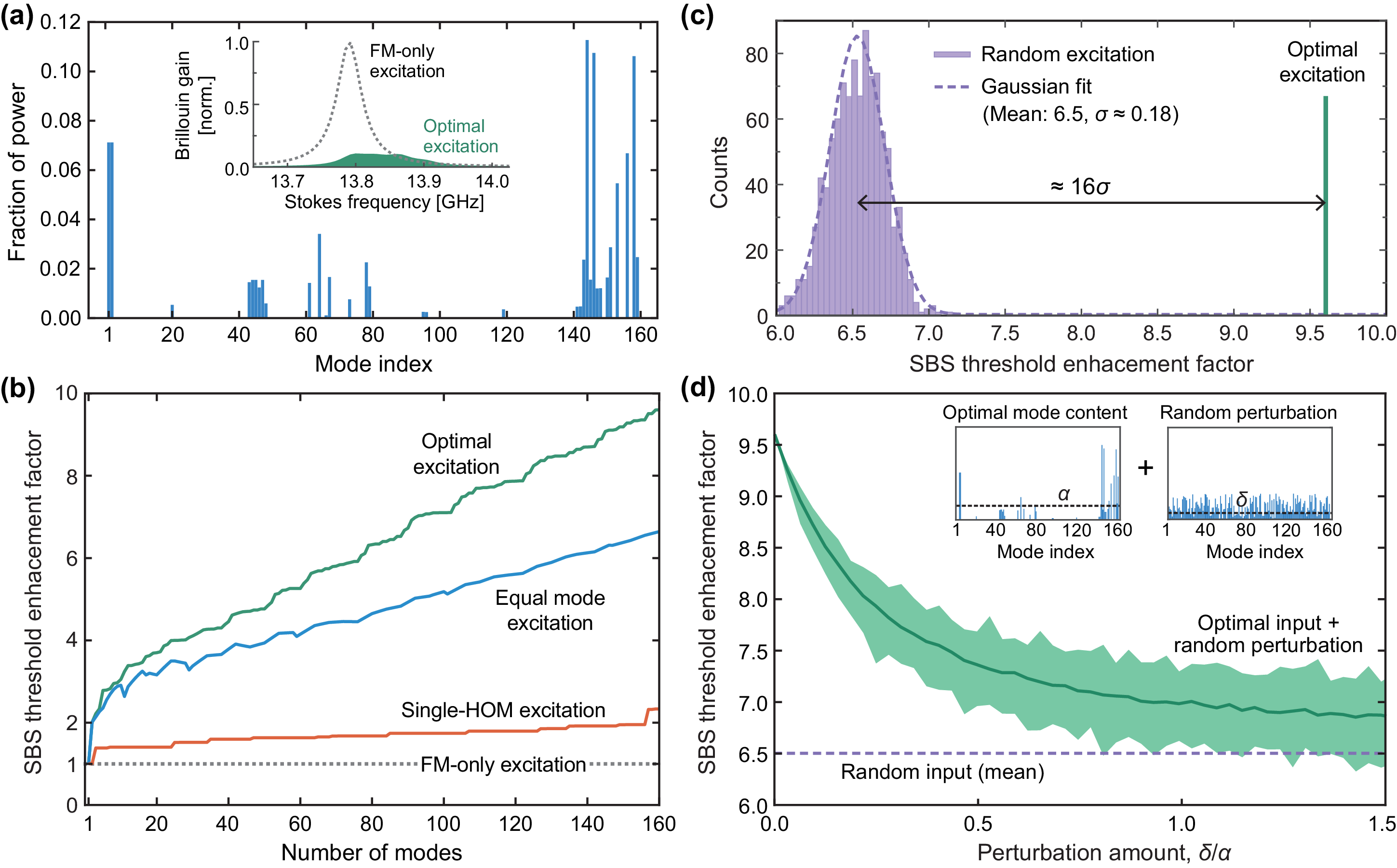}
    \caption{\textbf{Suppression of SBS with optimal input excitation in a step-index fiber amplifier with linear, mode-dependent gain.} (a)  Optimal mode content for 160 modes involves exciting a clusters of modes with a relatively larger weight to HOMs. This produces a significantly broadened Brillouin gain spectrum (inset) with much lower peak gain compared to FM-only excitation. (b) Scaling of SBS threshold enhancement (relative to FM-only excitation) with the number of modes for best single HOM excitation (orange), equal mode excitation (blue) and optimal excitation (green). The optimal excitation produces a significantly higher threshold enhancement than other two excitations with a maximum 9.6x enhancement. (c) A histogram of SBS threshold enhancement for 500 random input excitations. It follows a roughly Gaussian distribution with a mean value of 6.5x and the standard deviation $\sigma=0.18$. The optimal enhancement is roughly $16\sigma$ higher than the mean. (d) SBS threshold enhancement for optimal mode content with a random error of varying magnitude characterized by the standard deviation $\delta$. For scale, $\delta$ is compared with the standard deviation of the optimal mode content $\alpha$. The light-green region corresponds to various instances of random errors and the black curve shows the mean threshold enhancement for a fixed $\delta$. For small error magnitude the threshold enhancement is close to the optimal value and for large errors ($\delta>>\alpha$), it eventually converges to 6.5x. }
    \label{fig:Fig2}
\end{figure*}

We illustrate our general formalism for instability suppression by applying it first to the case of SBS suppression in a MMF. A semi-analytical theory of SBS for arbitrary multimode excitations was recently derived and experimentally validated in Refs.~\cite{wisal2022generalized,wisal2023theorySBS,chen2023mitigating}. It was shown that SBS results in the growth of a backward propagating Stokes wave (seeded by spontaneous Brillouin scattering) due to the scattering of the signal in various modes by acoustic phonons. The phonons are in turn generated by spatially varying optical forces created by the interference of signal and Stokes waves. The equation for the Stokes power growth in various modes can be obtained by solving coupled optical and acoustic wave equations, and is given by:

\begin{equation}
    \frac{dP^{\rm (S)}_m (\Omega,z)}{dz} = -\left[g(z)+\sum_l G_{\rm B}^{(m,l)} (\Omega) P_l(z)\right] P^{\rm (S)}_m(\Omega,z).
    \label{Eq:StokesEq}
\end{equation}
Here, $P^{\rm (S)}_m (\Omega,z)$ is the backward Stokes power in mode $m$ at Stokes frequency $\Omega$ at point $z$ along the fiber axis. $g(z)$ is the linear gain coefficient which is assumed to be mode-independent. The case of mode-dependent gain is discussed in detail in Sec.~IV. $G_{\rm B}^{(m,l)}$ is the Brillouin gain coefficient for mode pair $(m,l)$. $P_l(z)$ is the signal power in mode $l$, which is either constant (in passive low-loss fibers) or grows along the fiber axis (in fiber amplifiers) in $+z$ direction due to the stimulated emission. Equation~(\ref{Eq:StokesEq}) neglects mode coupling and polarization mixing due to fiber inhomogeneities, as well as loss in the passive fiber and mode-dependent gain and gain saturation in the active fiber, but captures the important physical features of multimode excitation.  The signal power can be treated as independent of the (initially very small) Stokes power, up to the SBS threshold (at which the Stokes power becomes a non-negligible fraction of the total power). It follows that the Stokes power in each mode $m$ grows exponentially in the backward direction and the power at the proximal end of the fiber is given by:
\begin{equation}
\begin{aligned}
    P^{\rm (S)}_m(\Omega,0)&=P^{\rm (S)}_m(\Omega,L) e^{\int_0^L dz g(z)} e^{\sum_l G_{\rm B}^{(m,l)} \int_0^L dz  P_l(z)}\\
    &= P^{\rm (S)}_m(\Omega,L) e^{g_{\rm av} L} e^{P_0L_{\rm eff}\sum_l G_{\rm B}^{(m,l)}\Tilde{P_l}}.
\end{aligned}
\label{Eq:StokeExp}
\end{equation}
Here, $L$ is the fiber length. $P^{\rm (S)}_m (\Omega,L)$ is the Stokes power in mode $m$ seeded by the noise at the distal end of the fiber. $P_0$ is the total output signal power, and we have defined $\Tilde{P}_l$ as the fraction of signal power in mode $l$ i.e., $\Tilde{P}_l=P_l (L)/P_0$. Assuming signal gain/loss is mode-independent, $\Tilde{P}_l$ is the same throughout the fiber. $g_{\rm av}$ is the averaged linear gain due to the stimulated emission and is given by $g_{\rm av}=\int_0^L dz\, g(z)/L $. The Stokes power in mode $m$ grows exponentially with growth rate given by a sum of the linear gain $g_{\rm av}$ and a nonlinear gain which is proportional to $P_0$ and a weighted sum of the Brillouin gain coefficients $G_{\rm B}^{(m,l)}$, with weights $\Tilde{P}_l$ depending on the input excitation. $L_{\rm eff}$ is the effective length of the fiber defined as $L_{\rm eff}=\frac{\int_0^L dz\, P_l(z)}{P_l(L)}$. For passive fibers with no loss, $L_{\rm eff}=L$, and in active fibers, $L_{\rm eff} < L$. In the absence of mode-dependent loss and gain, $L_{\rm eff}$ is the same for all the modes.

The Brillouin gain coefficient $G_{\rm B}^{(m,l)}$ represents the nonlinear gain in Stokes mode $m$ for a unit signal power in mode $l$. It can be calculated for any mode pair using an analytic formula which involves numerically evaluating the overlap integrals of relevant optical and acoustic modes. We consider a standard step-index fiber with 10-\textmu m core radius and numerical aperture (NA) of 0.3 supporting $160$ modes ($M=160$) including both polarizations. The Brillouin gain coefficients for all $160 \times 160$ mode pairs are calculated and stored at 100 different Stokes frequencies (in the range $[0,10]$ kHz). More details on the formula and the values of Brillouin gain coefficients are given in Supplement 1.

The SBS threshold~\cite{boyd2008nonlinear,kobyakov2010stimulated,wisal2023theorySBS} is typically set as the signal power level at which, for a given fiber length, the reflected Stokes power is $> 1\%$. From Eq.~\ref{Eq:StokeExp}, at frequency $\Omega_i$ each mode $m$ experiences growth which is exponential in $\sum_l G_{\rm B}^{(m,l)}(\Omega_i) \Tilde{P}_l$, and the SBS threshold will be determined by the single Stokes mode with the largest sum. Maximizing the SBS threshold, then, requires minimizing, over all possible input excitations $\Tilde{P}_l$, the maximum of the weighted Brillouin-gain-coefficient sums over all possible modes. Any distribution of input powers must satisfy two constraints: each mode weight is non-negative ($\Tilde{P}_l \geq 0$, and the total sum of weights equals one ($\sum_l \Tilde{P}_l = 1$.) Hence the input which maximizes the SBS threshold is the minimizer of the optimization problem:
\begin{equation}
    \min_{\{\Tilde{P}_l\}}\left[ \max_{m,\Omega_i} \sum_l G_{\rm B}^{(m,l)}(\Omega_i) \Tilde{P}_l\right],\:\:\:\: \sum_l \Tilde{P}_l = 1,\:\:\:\: \Tilde{P}_l \geq 0,\:\:\:\:l\in\{1,2,..M\},
    \label{Eq:SBSopt}
\end{equation}
where we consider $M$ modes for each of $N_{\Omega}$ Stokes frequencies. Both constraints of Eq.~\ref{Eq:SBSopt} are linear functions of the $P_l$ variables, while the maximum value of the weighted Brillouin coefficients is not. But there is a standard transformation that linearizes the problem: introduce a slack variable $t$ that is constrained to be larger than than all possible values of $\sum_l G_{\rm B}^{(m,l)(\Omega_i)} \Tilde{P}_l$, and minimize this variable. The discontinuities in the original objective are replaced by the intersection of $M \times N_{\Omega}$ linear constraints. We arrive at the transformed optimization problem:
\begin{equation}
\begin{aligned}
\min_{\{\Tilde{P}_l\},t} & t ,\\
\sum_l \Tilde{P}_l = & 1,\\
\Tilde{P}_l \geq 0,  \:\:\:\:& l\in\{1,2,..M\},\\
\sum_l G_{\rm B}^{(m,l)}(\Omega_i) \Tilde{P}_l \leq t, \:\:\:\:m\in\{1 &,2,..M\},  \:\:\:\:i\in\{1,2,..N_\Omega\}.
\end{aligned}
\label{Eq:SBSLP}
\end{equation}
This is a standard linear program with one equality constraint and ${(M+1)} N_{\Omega}$ inequality constraints. Linear programs are a subset of convex optimization problems, and globally optimal solutions can be computed with standard techniques (e.g., simplex or interior-point algorithms~\cite{kalai1997linear}). The inequality constraints define intersecting half-spaces and the equality constraint define a plane in $M+1$ dimensional real-euclidean space $\mathbb{R}^{M+1}$, which defines a convex polytope as the feasibility region for the solution. The optimal solution exists on the boundary of the feasibility region due to the convexity of linear functions~\cite{karloff2008linear,rigler2005optimization}.

We utilize this framework to find the optimal input excitation for maximal SBS threshold in the step-index MMF described above. We use the \emph{linprog} function in MATLAB~\cite{MATLAB}, which via the simplex algorithm can find an optimal solution for a 160-mode excitation on a MacBook Air laptop (1.6-GHz dual-core) in less than a minute. The optimal excitation leads to an SBS threshold 9.6$\times$ higher than the fundamental mode (FM)-only excitation in the same fiber (Fig.~\ref{fig:Fig2}b). 

The optimal mode content is shown in Fig.~\ref{fig:Fig2}a. The specific modal content is highly non-trivial and would be hard to predict from physical intuition only. However, the mode content does have certain qualitative features which can be understood in terms of the properties of the Brillouin gain coefficients $G_{\rm B}^{(m,l)}$. For instance, higher order modes (HOM) appear in a relatively higher fraction, since the Brillouin gain coefficients typically decrease with increasing mode order. Multiple modes are excited instead of a single HOM; this takes advantage of relatively lower intermodal gain coefficients ($l\neq m$) compared to the intramodal gain ($l=m$). Since this is the case, dividing power among many modes decreases the maximum modal gain. Finally, we observe that the optimal mode content involves a few clusters of modes with significant separation in the propagation constants, instead of exciting all the modes. This is because the Brillouin gain coefficients for mode pairs with significant difference between their propagation constant peak at very different frequencies. As a result, excitation of widely separated modes (as measured by their propagation constants) leads to a broadened Brillouin gain spectrum with a significantly lower peak gain value. This is highlighted in the inset of Fig.~\ref{fig:Fig2}a where we compare the Brillouin gain spectrum for FM-only excitation with the optimal excitation. This qualitative feature of widely separated modes was also observed in a recent experimental study on optimizing the SBS threshold via input control with a phase-only spatial light modulator~\cite{chen2023mitigating}. However, the maximum enhancement in SBS threshold achieved was lower than the globally optimal value, since the control over the input was limited and an iterative search algorithm was used to the find the optimal solution.

\begin{figure*}[t!]
    \centering
    \includegraphics[width=0.9\textwidth]{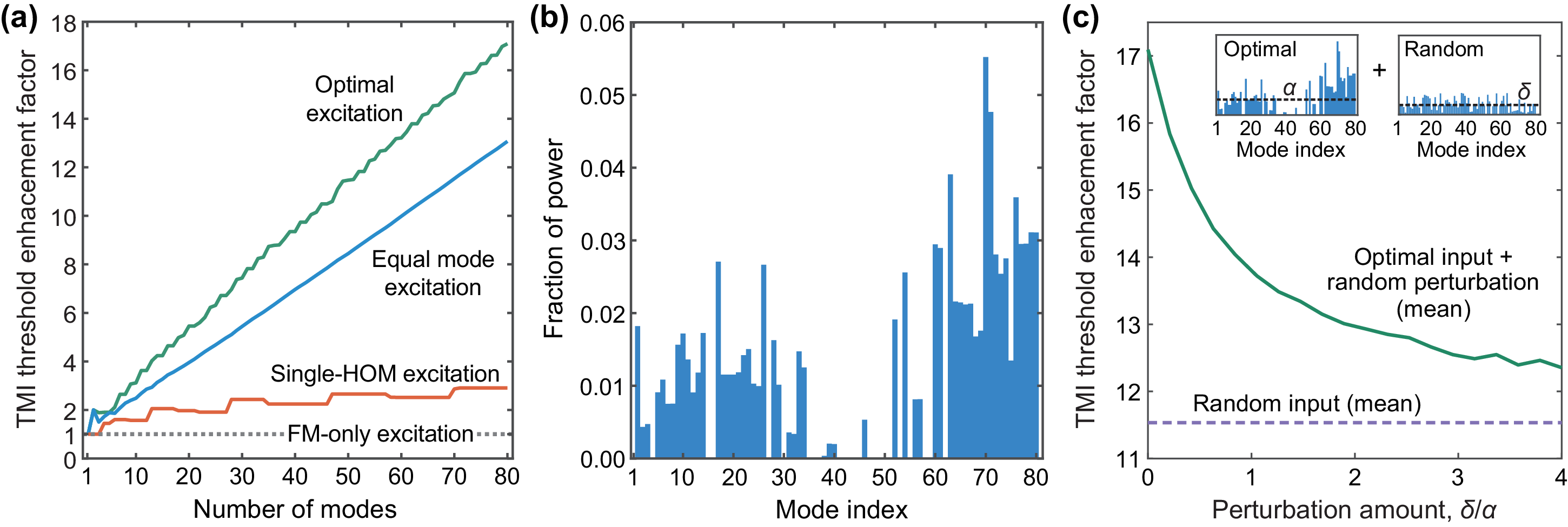}
    \caption{\textbf{Suppression of TMI with optimal input excitation in a step-index fiber amplifier with 82 modes} (a) Scaling of TMI threshold enhancement (relative to FM-only excitation) with the number of allowed modes for best single HOM excitation (orange), equal mode excitation (blue) and optimal excitation (green). The optimal excitation and equal-mode excitation leads to a linear increase in TMI threshold enhancement with maximum values of 17x and 13x respectively, much higher than single HOM excitation. (b) Optimal mode content involves excitation of almost all the modes with a relatively larger weight to highest order modes. A group of modes in the middle are not excited. These modes couple significantly with modes with relatively higher and lower order and not exciting them reduces the maximum thermo-optical coupling reducing the threshold. (c) TMI threshold enhancement for optimal mode content with a random error of varying magnitude (characterized by the standard deviation $\delta$). For scale, $\delta$ is compared with standard deviation of the optimal mode content $\alpha$. For small error magnitude the the threshold enhancement is close to the optimal value and eventually converges to 6.5x when $\delta>>\alpha$.}
    \label{fig:Fig3}
\end{figure*}

Next, we study the scaling of the SBS threshold enhancement with the number of modes in the fiber for the optimal excitation and compare it with two other types of input excitations, equal-mode excitation and best single-HOM excitation. The results are shown in Fig.~\ref{fig:Fig2}b. The SBS threshold enhancement is defined with respect to the FM-only excitation in all three cases. Single-HOM excitation shows minimal threshold enhancement, while equal-mode excitation does lead to a significantly higher threshold enhancement, which increases with the number of modes, reaching a maximum of 6.5$\times$ higher SBS threshold when all modes are excited. The enhancement of the SBS threshold upon optimal excitation also increases with the number of modes and is consistently higher than both the equal-mode and single-HOM excitations, reaching a maximum value of 9.6 with 160 modes. 

To understand the importance of finding a global optimum, we also study random input excitations with power in each mode chosen randomly from uniform $\rm [0,1]$ distribution with the total power normalized to unity. We calculate the SBS threshold enhancement for 500 such excitations and plot the histogram (shown in Fig.~\ref{fig:Fig2}c). The threshold enhancement factor over FM-only excitation roughly follows a Gaussian distribution with a mean value of 6.5 and the standard deviation $\sigma=0.18$, with most values falling between 6 and 7. The threshold enhancement factor upon optimal excitation is 9.6, which is 16$\sigma$ away from the mean. Assuming a normal distribution in the tail, this implies that 
a random search has a probability of $10^{-56}$ of finding the optimal mode content, showing the power of the optimization approach. 

Finally, we test the robustness of the optimal solution to small perturbations in the mode content. We start with the optimal mode content and add a noise with power randomly distributed and sampled from a uniform distribution with standard deviation $\delta$. We quantify the noise power by dividing $\delta$ with $\alpha$, the standard deviation of optimal-mode power distribution. The results for increasing noise are shown in Fig.~\ref{fig:Fig2}d. The light-green region shows the range of the threshold enhancement for various random instances, while the black curve represents the mean enhancement for a fixed $\delta$. A crucial observation (not visible in the plot) is that for small $\delta$ (where the excitation is close to our supposed global optimum) all of the values for the enhancement factor are smaller than 9.6, consistent with our claim of having found the global optimum.
As $\delta$ increases, the average threshold enhancement is close to the optimal value and decreases gradually as $\delta$ becomes large, instead of showing a sudden drop for arbitrarily small errors. This robustness is a consequence of the linearity of the optimization function and makes this method robust against experimental noise.  Eventually, when $\delta \gtrsim \alpha$, the mode content is essentially random and the mean threshold enhancement converges to 6.5, the mean enhancement for random input excitations. In the next section, we will show that this optimization approach can also be used to significantly suppress TMI.

\section{Optimal TMI Suppression}

TMI is a result of dynamic transfer of power between various optical modes due to nonlinear thermo-optical scattering~\cite{Cesar2020transverse} (see Fig.~\ref{fig:Fig1}c for the schematic). The equation for the noise power growth in various modes in a MMF can be obtained by solving coupled optical and heat equations, and were derived in Ref.~\cite{wisal2024theoryTMI}: 

\begin{equation}
    \frac{dP^{\rm (N)}_m (\Omega,z)}{dz} = \left[g+\sum_{l\neq m} \chi_{ml} (\Omega) P_l (z)\right] P^{\rm (N)}_m(\Omega,z).
\end{equation}
Here, $P^{\rm (N)}_m (\Omega,z)$ is the noise power in mode $m$ at a down-shifted frequency $\Omega$ at a distance $z$ along the fiber axis. $P_l(z)$ is the signal power in mode $l$. $\chi_{ml}(\Omega)$ is the thermo-optical coupling coefficient between mode pair $(m,l)$. Here we present the growth equations neglecting the effect that the signal power in each mode will grow in a different manner due to mode-dependent, non-linear gain saturation. We present the justification for neglecting this effect in our discussion section below.

Notice that the equation for noise power growth leading to TMI has the same form as the one for SBS (Eq.~\ref{Eq:StokesEq}), except for a few differences. Here, the negative sign on the right-hand side of the equation is absent, the Brillouin gain coefficients are replaced by the thermo-optical coupling coefficients and only the cross-couplings ($m \neq n$) are relevant. The thermal gratings responsible for the self-coupling terms ($m=n$) are much longer than the typical fiber length and therefore do not play a significant role. The thermo-optical coupling causes forward scattering and the noise power in each mode grows exponentially in the \emph{same} direction as the signal. As a result, for a large enough signal power (defined as the TMI threshold~\cite{Hansen2013theoretical,dong2013stimulated,wisal2024theoryTMI}), the noise in some mode can have a power equal to a significant fraction (typically set at $>1\%$) of the signal power and interfere with the signal, leading to a fluctuating output beam profile. Similar to the case of SBS, the condition for finding the optimal input excitation leading to maximum TMI threshold is given by:

\begin{equation}
    \min_{\{\Tilde{P}_l\}}\left[ \max_{m,\Omega_i} \sum_{l\neq m} \chi_{ml}(\Omega_i) \Tilde{P}_l\right],\:\:\:\: \sum_l \Tilde{P}_l = 1,\:\:\:\: \Tilde{P}_l \geq 0,\:\:\:\:l\in\{1,2,..M\}.
    \label{Eq:TMIopt}
\end{equation}
\noindent
Here, $\Tilde{P}_l$ is the fraction of signal power in mode $l$. The optimization involves finding a distribution of signal power $\{\Tilde{P}_l\}$ in various modes that minimizes the maximum noise growth rate (proportional to the weighted sum $\sum_{l\neq m} \chi_{ml}(\Omega_i) \Tilde{P}_l$) over all the noise modes $m$ and frequencies $\Omega_i$. By the same slack-variable technique Eq.~\ref{Eq:TMIopt} can be transformed into a standard linear program similar to Eq.~\ref{Eq:SBSLP}, whose global optima can be obtained by any linear-programming solver.

We consider a step-index MMF amplifier with a core radius of $20$ \textmu m and NA$=0.15$ which supports 80 modes per polarization at $\lambda=1064$~nm. First, we calculate $\chi_{ml}$ for all the mode pairs at 100 frequencies (in the range $[0,10]$ kHz) using an analytical formula involving the overlap integrals of relevant optical and thermal modes. More details on the formula used and the resulting values of $\chi_{ml}$ are provided in the supplementary. We find that $\chi_{ml}$ is a highly sparse matrix and leads to strong coupling only between the optical modes with similar transverse spatial frequencies.  This is a result of the diffusive nature of underlying heat propagation which exponentially dampens high-spatial-frequency features in the temperature fluctuations. The sparse coupling matrix generically favors multimode excitation for achieving a lower effective thermo-optical coupling.

We calculate the optimal excitation and the associated enhancement in TMI threshold for a variable number of excited modes in the fiber. The optimal mode content for when all 82 modes are considered are shown in Fig.~\ref{fig:Fig3}a. Generically most modes are excited with relatively higher weight given to higher order modes. These features in the optimal mode content are consistent with the sparse nature of $\chi_{ml}$ which favors multimode excitation and relatively lower value of $\chi_{ml}$ for high mode order which favors exciting the HOMs. Surprisingly, the optimal mode content reveals a new strategy to further increase the threshold --- \emph{not exciting a group of modes in the middle} (i.e., with intermediate propagation constants). The modes in the middle have a significant number of `neighboring modes' with both higher and lower mode indices. As a result, these modes experience the highest thermo-optical coupling, and avoiding these modes increases the TMI threshold. This displays the power of the optimization approach, which not only reveals the the maximum level of threshold enhancement possible upon multimode excitation but also deepens our understanding of the strategies to achieve the maximum enhancement. For comparison, we also calculate the scaling of the TMI threshold enhancement for equal-mode excitation and single-HOM excitation and study the scaling with the number of modes considered. The results are shown in Fig.~\ref{fig:Fig3}b. In each case, the TMI threshold enhancement is defined relative to the FM-only excitation. Both optimal and equal-mode excitation perform significantly better than both FM-only and single-HOM excitations and lead to a linear increase in the TMI threshold with the number of allowed fiber modes. This is consistent with the sparse nature of thermo-optical coupling, mentioned earlier. The slope of the linear scaling is significantly higher for the optimal excitation (0.20) compared to the equal-mode excitation (0.15). When all 82 modes are excited, the optimal excitation and equal-mode excitation lead to 17$\times$ and 13$\times$ higher TMI thresholds than that of the FM-only excitation, respectively. 

Finally, we test the robustness of the optimal solution to small perturbations in the mode content. We follow the same procedure as we did for SBS in the previous section. We add a randomly chosen mode content of varying relative magnitude to the optimal mode content and calculate the TMI threshold enhancement. The results are shown in Fig.~\ref{fig:Fig3}c.  For small errors, the threshold enhancement is close to the optimal value and decreases slowly as the error becomes large, instead of a sharp drop for arbitrarily small errors. Similar to SBS, this robustness is a consequence of linearity of the optimization function and increases the experimental feasibility of this method.  

\section{Impact of Mode-dependent gain and loss}

In the analysis so far, for simplicity, we have assumed that the signal power in all the modes experience the same amount of linear gain (in active fibers) or loss (in passive fibers). However, in real experiments this is not strictly true. In passive fibers, the signal experiences loss due to various fiber imperfections, which is typically higher for higher order modes~\cite{ho2011statistics}. Also, in active fibers, the fundamental mode experiences the highest linear gain and on average the gain coefficient decreases with the mode order~\cite{bai2011multimode}. In this section, we incorporate both of these effects in our optimization formalism and study their impact on the optimal mode content for SBS suppression and the SBS threshold enhancement factor. Similar analysis can be performed for the TMI suppression as well but is not presented here. We combine mode-dependent gain and loss to define a relative MDL. To quantify the relative MDL, we define a `loss' coefficient for each HOM relative to the FM:

\begin{figure}[t!]
    \centering
    \includegraphics[width=0.45\textwidth]{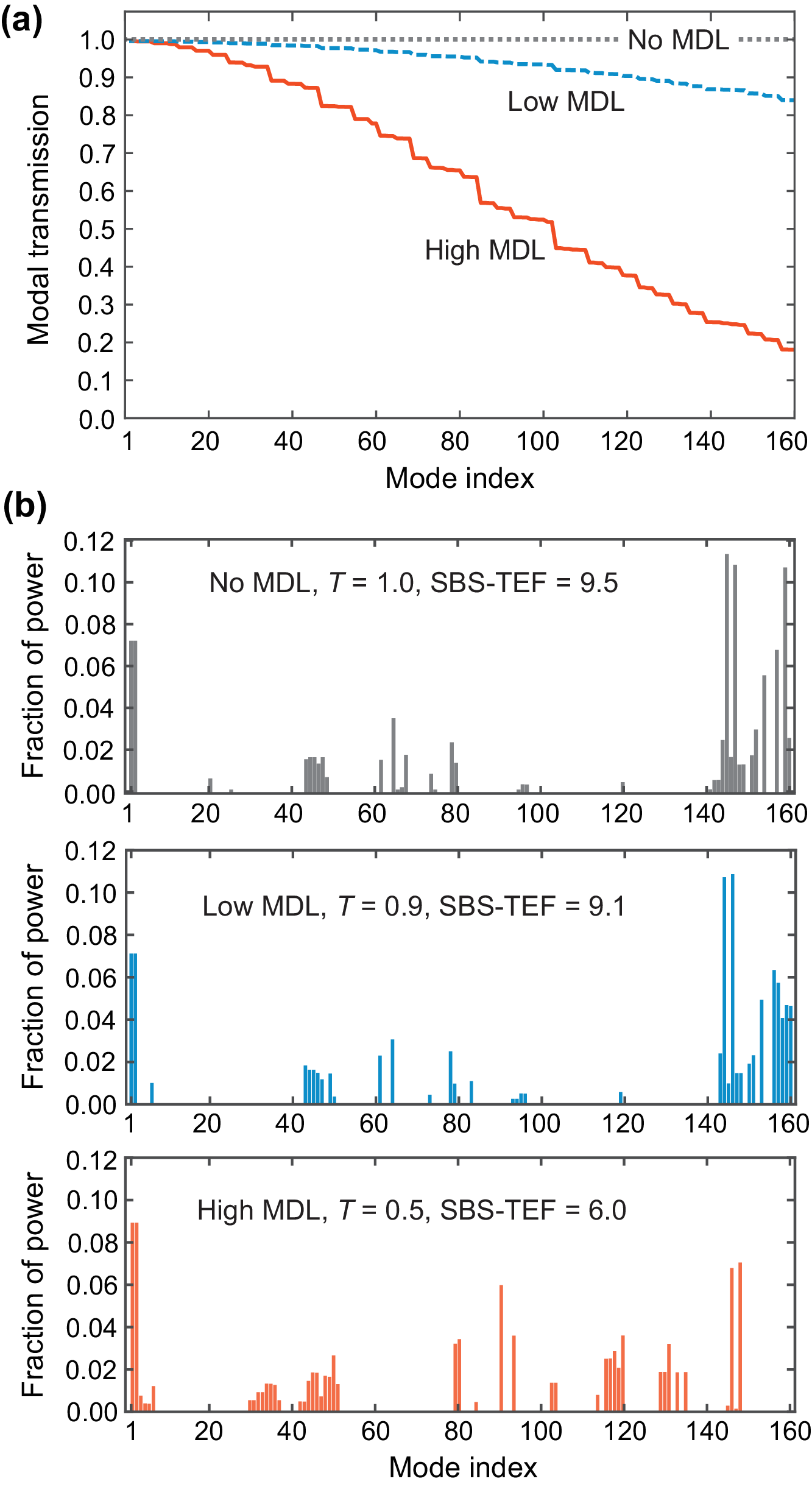}
    \caption{\textbf{Impact of mode-dependent loss (MDL) on SBS suppression upon optimal excitation} (a) Relative modal transmission for three different levels of MDL: no MDL (dotted curve), low MDL (dashed curve), and high MDL (solid curve). (b) Optimal mode content for SBS suppression in fibers with different levels of MDL mentioned in (a). When the MDL is relatively low, the optimal mode content looks very similar to the case of no MDL. The SBS threshold enhancement factor (SBS-TEF) reduces slightly from 9.6 to 9.1 and the net transmission is 0.9. When the MDL is higher, the optimal mode content changes substantially with increased weight to modes in the middle and reduced weight to the highest order modes. This is the result of higher effective length for the modes with higher loss coefficient leading to increased Stokes gain when these mode are excited. }
    \label{fig:Fig4}
\end{figure}

\begin{equation}
    \alpha_l = g_l - g_1,
\end{equation}
Here, $\alpha_l$ is the relative loss coefficient and $g_l$ is the net linear gain coefficient for mode $l$. Negative values of $g_l$ mode $l$ has net loss. In general, $\alpha_l$ is negative, and it's magnitude increases with mode order. In active fibers, $g_1>g_l>0$, whereas in lossy fibers $g_1\approx0,\: g_l<0$. We define a relative transmission of each mode $T_l = e^{\alpha_l L }$. By definition the transmission of the fundamental mode is normalized to unity ($T_1=1$) and $T_l$ decreases with mode index $l$.

The main effect of relative MDL is that the effective length $L_{\rm eff}$, which determines the Stokes power growth (see Eq.~\ref{Eq:StokeExp}), becomes mode dependent. Recall that, we defined $L^{(l)}_{\rm eff} = {\int_0^L P_l(z)}/{P_l(L)}$. Relative MDL leads to a variable growth rate for signal power in different modes, hence $L_{\rm eff}$ depends on mode index $l$. For an active fiber, assuming the power in mode $l$ grows with gain coefficient $g_l$:

\begin{equation}
    L^{(l)}_{\rm eff} = \frac{e^{g_{l}L} -1}{g_l e^{g_{l}L}}\approx  \frac{1}{g_l}.
    \label{Eq:Lgain}
\end{equation}
For significant amplification, the effective length is inversely proportional to the gain coefficient and increases with mode order. For passive fibers with no gain and a relative loss coefficient $\alpha_l$ in mode $l$:

\begin{equation}
    L^{(l)}_{\rm eff} = \frac{e^{-|\alpha_{l}|L}-1}{-|\alpha_l|e^{-|\alpha_{l}|L}} =L (\frac{e^{|\alpha_{l}|L}-1}{{|\alpha_l| L}})\approx L (1+|\alpha| L/2).
\end{equation}
The factor multiplying $L$ in the last term increases with $|\alpha_l|$ and is equal to unity for $\alpha=0$. Therefore, $L^{(l)}_{\rm eff}$ increases with mode order for passive fibers too.  The mode dependence of the effective length modifies the optimization condition (Eq.~\ref{Eq:SBSopt}) as follows:

\begin{equation}
    \min_{\{\Tilde{P}_l\}}\left[ \max_{m,\Omega_i} \sum_l L^{l}_{\rm eff} G_{\rm B}^{(m,l)}(\Omega_i) \Tilde{P}_l\right],\:\:\:\: \sum_l \Tilde{P}_l   = 1,\:\:\:\: \Tilde{P}_l \geq 0,\:\:\:\:l\in\{1,2,..M\}.
    \label{Eq:SBSoptMDL}
\end{equation}
The main modification in the objective function is the inclusion of $L^{(l)}_{\rm eff}$ in the argument of the sum. Since $L^{(l)}_{\rm eff}$ increases with $l$, this effectively increases the Stokes gain for signal power in higher-order modes. Physically, since the SBS threshold is defined for the output power, higher signal loss means more stokes gain for fixed output power.  As a result, we expect that the presence of relative MDL will lead to a decrease in the contribution of HOMs in the optimal solution.

To quantitatively study the impact of MDL, we consider the passive step-index MMF discussed in Sec.~II with three different levels of loss: (1) no MDL (2) low MDL and (3) high MDL. The modal transmission for the three cases is shown in Fig.~\ref{fig:Fig4}a with curves in different colors. The relative loss coefficient of a mode is assumed to vary quadratically with the mode index, and the proportionality constant is varied to achieve different levels of loss. This phenomenological model of MDL can be derived by \textit{ab-initio} methods considering the fiber bending/twisting and the scattering due to disorder and has been verified experimentally~\cite{mickelson1983mode,olshansky1976mode,ho2012exact,chen2023mitigating}. For the three cases of MDL, we find the optimal mode content and the associated SBS threshold enhancement factor. The results are shown in Fig.~\ref{fig:Fig4}b, with different colors corresponding to different loss levels. When the MDL is relatively low, the optimal mode content looks very similar to the case of no MDL. The threshold enhancement factor reduces slightly from 9.6 to 9.1 and the net transmission is 0.9. When the MDL is higher, the optimal mode content changes substantially with increased weight to modes in the middle and reduced weight to the highest-order modes. This is the result of higher effective length for the modes with higher loss coefficient leading to increased Stokes gain when these modes are excited. The optimal SBS threshold enhancement is roughly 6$\times$, which is still higher than the enhancement for equal-mode excitation (4.9$\times$) and much larger compared to the FM-only excitation. Note that we define the SBS threshold for the output signal power, therefore the 6$\times$ enhancement here already takes into account the decrease in the output power at the SBS threshold due to lowered transmission as a result of the loss. The transmission in this case is 0.5 which necessitates a higher input power to reach the maximum SBS threshold. In a fiber amplifier, presence of MDL will lead to reduced efficiency upon HOM excitation but will increase the maximum reachable output power before SBS becomes prohibitive. These results confirm that our optimization formalism is useful even in the presence of mode-dependent gain or loss and the optimal multimode excitation can still lead to substantial enhancement to instability-limited output power compared to FM-only excitation.

\section{Discussion and conclusion}

In this work, we present an approach that provides the optimal input excitation for maximal output power thresholds for stable operation in any MMF, limited by two nonlinear instabilities, SBS and TMI. The key insight involved transforming both SBS and TMI suppression into standard linear programs, allowing us to find globally optimal solutions via existing numerical solvers. The optimal excitations lead to substantially higher output power limited by SBS and TMI, displaying the power of the optimization approach. 

In our formalism, we are able to take into account mode-dependent loss and gain~\cite{ho2011statistics,ho2012exact,mickelson1983mode,bai2011multimode}.  The optimization approach results in significant SBS threshold enhancement even with high MDL. For simplicity in our analysis, we considered the signal gain to be linear and ignored any gain saturation. This assumption can break down in high-power amplifiers where gain saturation can be a significant effect~\cite{paschotta1997ytterbium,Smith2013increasing,hansen2014impact}. If needed, this assumption can possibly be relaxed in our formalism. The key modification would be that $L^{(l)}_{\rm eff}$ would need to be computed numerically, instead of using a simple analytic formula as we did in Eq.~\ref{Eq:Lgain}. For SBS, gain saturation would not have a significant impact on the optimal excitation and the optimal threshold enhancement, except for the increased mode-dependent gain via spatial-hole burning~\cite{jiang2008impact}. For TMI, the gain saturation would play a more significant role as the heat load is also impacted by the gain saturation, which modifies the thermo-optical coupling~\cite{Smith2013increasing,hansen2014impact, dong2022accurate}. The qualitative characteristics of the optimal excitation are still expected to remain unchanged since the qualitative nature of the thermo-optical coupling is not modified on average.

Experimental implementation of our approach can be achieved by selective mode excitation with an SLM. Nearly exact modal distribution can be achieved via both amplitude and phase modulation in an SLM, which has been demonstrated~\cite{gomes2022near}. In recent experiments, it was demonstrated that SBS threshold can be improved substantially via input wavefront shaping in a standard passive MMF. However, only phase-modulation of the SLM was considered, leading to limited control, and the globally maximal threshold enhancement was likely not reached~\cite{chen2023mitigating}. Our theory assumes an ideal fiber, with no random linear mode coupling~\cite{ho2011statistics}. This is typically valid if the fiber is of high quality and not coiled very tightly, and the fiber is not too long. For weak mode coupling the theoretically predicted optimal solution should produce instability threshold enhancement factor close to the optimal value, as shown by the robustness to weak perturbations in Fig.~\ref{fig:Fig2}d and Fig.~\ref{fig:Fig3}c. For fibers with strong random linear mode coupling, the theoretically optimal solution will no longer be correct. However, even in such cases an optimal solution can be found via search algorithms~\cite{tzang2018adaptive,florentin2017shaping}, although finding the globally optimal solution is not guaranteed. Even for these cases, our method can be helpful as it provides an upper bound on the maximum possible enhancement in SBS or TMI threshold via multimode excitations. Additionally, the physical insights gained into the SBS suppression strategy from the optimal mode content can be utilized in improving the search algorithms in these experiments.

In this work, we focused on increasing the output power limited by SBS and TMI, while ignoring any other nonlinear instabilities. Input wavefront shaping in MMFs for controlling stimulated Raman scattering and four-wave mixing due to Kerr nonlinearity has been demonstrated in passive fibers~\cite{tzang2018adaptive}.  The input excitations in these studies were obtained with feedback optimization via a genetic algorithm. As such, the globally optimal solution was likely not obtained. Our optimization approach for finding the globally optimal solution can possibly be extended to these nonlinearities as well for both passive and active fibers. The key requirement to apply our formalism is the linearity of the instability gain in terms of the control parameters, along with any constraints. 

Wavefront shaping provides a novel and exciting tool for controlling nonlinear phenomena. Our work contributes to this emerging field by providing a new tool to find optimal input wavefront for nonlinear instabilities, helping address important practical challenge in achieving ultra-high laser powers.

\section*{acknowledgments}

We thank Ori Henderson-Sapir, Heike Ebendorff-Heidepriem, and David Ottaway at The University of Adelaide, Stephen Warren-Smith and Linh Viet Nguyen at University of South Australia, and Peyman Ahmadi at Coherent for stimulating discussions. We acknowledge the computational resources provided by the Yale High Performance Computing Cluster (Yale HPC). This work is supported by the Air Force Office of Scientific Research (AFOSR) under Grant FA9550-20-1-0129. We also acknowledge the support of Simons Collaboration on Extreme Wave Phenomena Based on Symmetries.

\section*{author declaration}

The authors declare no conflict of interest.

\section*{Data availability}

Data underlying the results presented in this paper are not publicly available at this time but may be obtained from the authors upon reasonable request.

\newpage
\textbf{\huge Supplementary Information}

 \renewcommand{\theequation}{S.\arabic{equation}}
 \setcounter{equation}{0} 
 \renewcommand{\thesection}{S.\arabic{section}}
 \setcounter{section}{0} 
 
\renewcommand\thefigure{S.\arabic{figure}} 
  \setcounter{figure}{0}
  
	\section{Multimode SBS Suppression}

		\begin{figure*}[b!]
			\centering
			\includegraphics[width=0.8\textwidth]{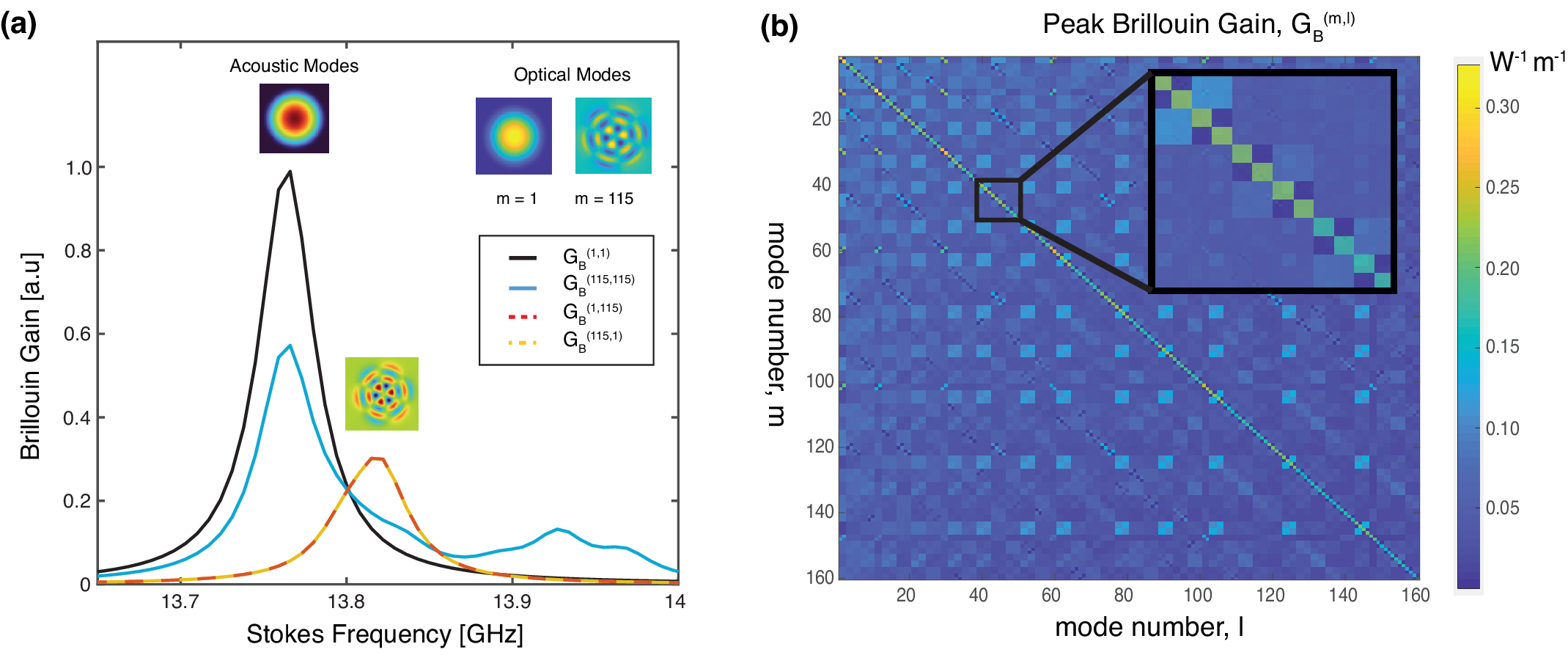}
		
		\caption{ \textbf{Brillouin gain coefficients in multimode fibers.} (a) Intramodal and intermodal gain for the fundamental mode (m=1) and a higher mode mode (m=115) in a circular step-index fiber. Optical mode profiles are shown in the inset. Intermodal gain has relatively lower peak value and peaks at a higher Stokes frequency as it has major contribution from higher order acoustic modes (shown above each curve). (b) Peak value of Brillouin gain for all the mode pairs.  Off diagonal entries (intermodal gain) have relatively lower value compared to the diagonal entries (intramodal gain).  }     
		\label{fig:SBSgain}
            \end{figure*}
            
	\subsection{Brillouin gain coefficients}

In section II of the main text, we find the optimal input mode content for maximum SBS threshold in a realistic multimode fiber. A key step in this process is calculating the Brillouin gain coefficients $G^{(l,m)}_{\rm B}(\Omega)$ for various optical mode pairs $\{l,m\}$ and Stokes frequencies $\Omega$ for the given multimode fiber. For this calculation, we utilize the formula for multimode Brillouin gain coefficients derived in Ref.~\cite{wisal2023theorySBS}, which is given by:

\begin{equation}
    G^{(l,m)}_{\rm B}(\Omega) =G_0\sum_{k}{|\langle\vec{\psi}_m^*\cdot \vec{\psi}_l u_k\rangle|}^2\frac{\frac{\Gamma_{mlk}}{2}}{(\Omega_{mlk}-\Omega)^2+{(\frac{\Gamma_{mlk}}{2})}^2}. 
    \label{Eq:Brillouin}
\end{equation}

The Brillouin gain coefficient for optical mode pair $(\vec{\psi}_l,\vec{\psi}_m^*)$ at Stokes frequency $\Omega$ is a
sum of Lorentzian curves for each acoustic mode $u_k$ with
a center frequency equal to the acoustic eigenfrequency
$\Omega_{mlk}$, and a linewidth equal to the effective acoustic loss
$\Gamma_{mlk}$. Each acoustic mode contribution is proportional to 
the corresponding integral over the fiber cross-section (denoted by angle brackets) of the dot product of the optical
mode profiles ($\vec{\psi}_m$ and $\vec{\psi}_l$) multiplied with the scalar acoustic mode profile $u_k$. $G^{(l,m)}_{\rm B}$ is also proportional to a constant $G_0$, which depends on various optical and acoustic properties of the fiber and is given by $G_0\approx\frac{2\gamma_{\rm e}^2\omega_{\rm s}^2}{n_{\rm c}v_{\rm a} c^3}$. Here, $\gamma_{\rm e}$ is the electrostriction constant, $\omega_{\rm s}$ is the optical signal frequency, $n_{\rm c}$ is the core refractive index, $v_{\rm a}$ is the core acoustic velocity and $c$ is the speed of light in vacuum.

            \begin{figure}[h!]
			\centering
			\includegraphics[width=0.45\textwidth]{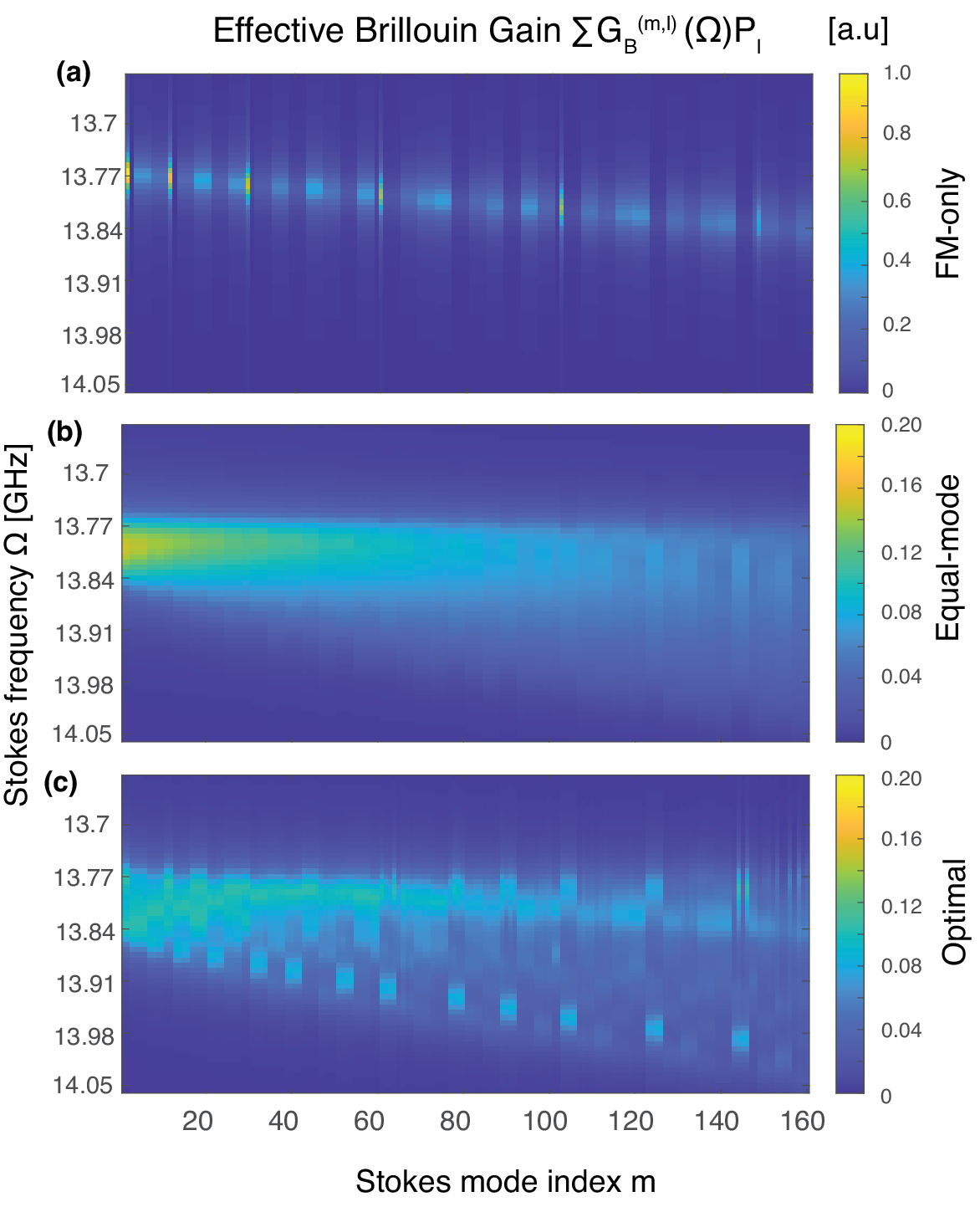}
		
		\caption{ \textbf{ Effective Brillouin gain in various Stokes modes and frequencies 
 for three different input excitations} (a) For fundamental mode (FM)-only excitation, effective Brillouin gain is concentrated in the fundamental Stokes mode at around 13.77 GHz, with relatively large peak gain, leading to a low SBS threshold. A few other radial modes which are azimuthally symmetric also have significant gain.  (b) For equal mode excitation, effective Brillouin gain spreads over multiple Stokes modes and broadens in frequency range leading to a significant reduction in the peak value (notice the change in the colorbar scale) and a 6.5$\times$ higher SBS threshold. (c) For optimal excitation, gain broadening over Stokes modes and frequencies is maximal with much lower peak gain, leading to 9.5$\times$ higher SBS threshold. The sum of Brillouin gain across all Stokes modes and frequencies is roughly constant for different excitations.}        
		\label{fig:heatmap}
            \end{figure}

We utilize the above formula to calculate the Brillouin gain coefficients in a 160-mode step-index fiber with a core radius of 10 \textmu m and the numerical aperture (NA) of 0.3, for all the mode pairs and multiple Stokes frequencies in the range of 13.5 GHz to 14.5 GHz. For $l=m$ we refer to the gain as intramodal gain and for $l\neq m$ we refer to it as intermodal gain. To illustrate some generic properties of $G^{(l,m)}_B(\Omega)$, we plot the two intramodal and intermodal curves for the fundamental mode (FM, $m=1$) and a higher-order mode (HOM, $m=120$) in Fig.~\ref{fig:SBSgain}a. These figures are similar to Fig.~3a and 4a in Ref.~\cite{wisal2023theorySBS} and are plotted here for the convenience of the readers. The relevant optical mode profiles are shown in the inset in the top right of the figure. The intramodal gain for the FM (black curve) has the highest peak value followed by the intramodal gain for the HOM (blue curve). The two intermodal gain curves (red and yellow) are identical and have much lower peak values compared to the intramodal gain. The intermodal gain curves also peak at a relatively higher Stokes frequency as their main contribution comes from a higher order acoustic mode (shown in the inset) with higher acoustic eigenfrequency, whereas intramodal gain (for both modes) has its major contribution from the fundamental acoustic mode. These properties hold generically for most optical mode pairs. Peak values of Brillouin gain for all the mode pairs is shown in Fig.~\ref{fig:SBSgain}b as a false color matrix. The off-diagonal entries (intermodal gain) have relatively lower value compared to the diagonal entries (intramodal gain), confirming the above result.
The relatively weaker and spectrally shifted intermodal gain favors multimode excitation for SBS suppression over FM-only or single-HOM excitation. 

\subsection{Effective SBS gain and gain broadening}

For a given signal power distribution $\{\Tilde{P}_l\}$, the effective Brillouin gain in Stokes mode $m$ at Stokes frequency $\Omega_i$ is given by $\sum_l G^{(l,m)}_{\rm B}(\Omega_i) \Tilde{P}_l$. The overall SBS gain is given by the maximum gain across all the Stokes modes and frequencies, and it is inversely proportional to the SBS threshold. In Fig.~\ref{fig:heatmap}, we plot the effective Brillouin gain in all the Stokes modes at varying Stokes frequencies for three different input excitations (1) FM-only (2) equal-mode and (3) optimal excitation. For the FM-only excitation the effective Brillouin gain is concentrated in the fundamental mode and in a narrow frequency range. As such the maximum value of Brillouin gain is relatively high, leading to a low SBS threshold. A few other radial modes which are azimuthally symmetric also have significant gain. For equal mode excitation, the effective Brillouin gain spreads over multiple modes and a larger frequency range leading to significantly reduced peak gain, increasing the SBS threshold by a factor of 6.5 compared to the FM-only excitation. The optimal excitation leads to an even higher spreading of gain across different Stokes frequencies and modes, increasing the SBS threshold by a factor of 9.5. It can be seen that for both cases of multimode excitations (equal-mode and optimal) effective brillouin gain extends towards higher Stokes frequencies, which is a result of relative blue-shifting of intermodal SBS gain which has high contribution in multimode excitations. For the optimal excitation, some modes have doubly peaked effective Brillouin gain. This happens because optimal excitation consists of exciting clusters of modes (separated in momentum-space) which produce spectrally separated intramodal and intermodal gain contributions in any mode, and the sum of these contributions lead to a generically broadened and for some modes doubly peaked gain spectra. Note that the total Brillouin gain $\sum_{m,\Omega_i}\sum_l G^{(l,m)}_B(\Omega_i) \Tilde{P}_l$ is roughly constant for different input excitations ($ <10 \% $ variation between the three cases). This allows multimode excitations to increase the SBS threshold by decreasing the peak gain via spreading of gain across multiple modes and frequencies.

\section{Multimode TMI Suppression}

	\subsection{Thermo-optical coupling coefficients}

\begin{figure}[t!]
			\centering
			\includegraphics[width=0.45\textwidth]{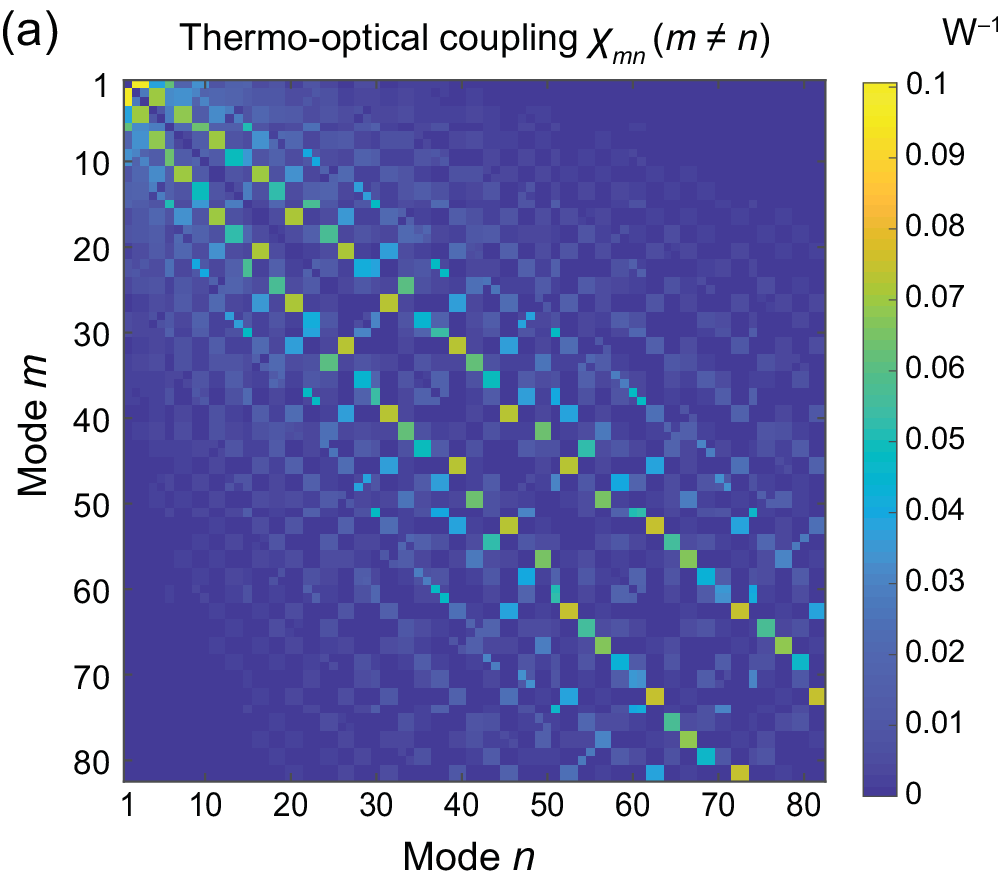}
		
		\caption{ \textbf{Thermo-optical coupling coefficients for all the mode pairs in a 82-mode fiber amplifier.} Only cross coupling ($m\neq n$) is considered, since it is most relevant for TMI. The coupling matrix is tightly banded and highly sparse with only a few entries near the diagonal having large values. This is a result of weak coupling between modes with large spatial-frequency mismatch due to the exponential decay of high spatial frequency temperature fluctuations needed to couple such mode pairs. }        
		\label{fig:TMIgain}
            \end{figure}

In section III of the main text, we find the optimal input mode content for maximum TMI threshold in a realistic multimode fiber amplifier. Similar to SBS suppression, a key step in this process is calculating the thermo-optical coupling coefficients $\chi_{ml}(\Omega)$ for various optical mode pairs $\{l,m\}$ and Stokes frequencies $\Omega$ for the given multimode fiber. For this calculation, we utilize the formula for multimode thermo-optical coupling coefficients derived in Ref.~\cite{wisal2024theoryTMI}, which is given by:

\begin{equation}
    \chi_{ml}(\Omega) =\chi_0\sum_{k}{|\langle\vec{\psi}_m^*\cdot \vec{\psi}_l \Tilde{T}_k\rangle|}^2\frac{D\Omega}{\Omega^2+{\Gamma_k}^2}. 
    \label{Eq:Brillouin}
\end{equation}
 Here $D$ is the diffusion constant and $\chi_0$ is an overall coupling constant which depends on various thermal and optical properties of the fiber and is given by $\chi_0={\eta q_{\rm D} k_0}/{2 n_{\rm c} \kappa}$. $\eta$ is the thermo-optic coefficient, $q_{\rm D}$ is the size of the quantum defect, $k_0$ is the free-space wavenumber, $n_{\rm c}$ is the linear refractive index of the fiber core, and $\kappa$ is the thermal conductivity of the fiber.  The thermo-optical coupling coefficient between any two modes $\chi_{ml}$ has contributions from each temperature eigenmode $\Tilde{T}_k$. The strength of each contribution is proportional to the integrated overlap of the dot product of the optical mode profiles $\vec{\psi}_m$ and $\vec{\psi}_l$ multiplied with the temperature mode profile $\Tilde{T}_k$. Each contribution has a characteristic frequency curve which peaks at a frequency given by inverse mode lifetime $\Gamma_k$, and the peak value is proportional to the thermal mode lifetime $4\pi\Gamma_k^{-1}$. Higher-order temperature eigenmodes (with high spatial frequencies) decay quickly and hence have lower mode lifetime, characteristic of the diffusive nature of heat propagation. As a result the coupling between optical modes with large transverse spatial-frequency mismatch (facilitated by higher-order temperature modes) is expected to be quite weak. We utilize the above formula to calculate thermo-optical coupling coefficients between all the mode pairs in a circular step index fiber with core radius of 20 \textmu m and NA of $0.15$ supporting 82 modes per polarization. In Fig.~\ref{fig:TMIgain} we have plotted the peak values of thermo-optical coupling coefficient between all the mode pairs as a false color matrix. This figure is similar to Fig.~4a in Ref.~\cite{wisal2024theoryTMI} and is plotted here for the convenience of the readers. The resultant coupling matrix is highly sparse and tightly banded with strong coupling only between the mode pairs with similar spatial frequencies. Such a coupling matrix generically favors multimode excitation over both fundamental mode-only or single higher order mode excitation. Our linear-programming-based optimization approach optimally utilizes the sparse structure of the coupling matrix to obtain the multimode excitation corresponding to  the minimum TMI gain and maximum TMI threshold.

\end{document}